# Understanding Dynamics in Coarse-Grained Models: I. Universal Excess Entropy Scaling Relationship

Jaehyeok Jin,[1] Kenneth S. Schweizer,[2] and Gregory A. Voth[1*]

[1] Department of Chemistry, Chicago Center for Theoretical Chemistry, Institute for Biophysical Dynamics, and James Franck Institute, The University of Chicago, Chicago, IL 60637, USA

[2] Department of Material Science, Department of Chemistry, Department of Chemical & Biomolecular Engineering, and Materials Research Laboratory, University of Illinois, Urbana, IL 61801, USA

* Corresponding author: gavoth@uchicago.edu

**Abstract**
Coarse-grained (CG) models facilitate an efficient exploration of complex systems by reducing the unnecessary degrees of freedom of the fine-grained (FG) system while recapitulating major structural correlations. Unlike structural properties, assessing dynamic properties in CG modeling is often unfeasible due to the accelerated dynamics of the CG models, which allows for more efficient structural sampling. Therefore, the ultimate goal of the present series of articles is to establish a better correspondence between the FG and CG dynamics. To assess and compare dynamical properties in the FG and the corresponding CG models, we utilize the excess entropy scaling relationship. For Paper I of this series, we provide evidence that the FG and the corresponding CG counterpart follow the same universal scaling relationship. By carefully reviewing and examining the literature, we develop a new theory to calculate excess entropies for the FG and CG systems while accounting for entropy representability. We demonstrate that the excess entropy scaling idea can be readily applied to liquid water and methanol systems at both the FG and CG resolutions. For both liquids, we reveal that the scaling exponents remain unchanged from the coarse-graining process, indicating that the scaling behavior is universal for the same underlying molecular systems. Combining this finding with the concept of mapping entropy in CG models, we show that the missing entropy plays an important role in accelerating the CG dynamics.

# I. Introduction

With the aid of computation, coarse-grained (CG) models allow for efficient exploration of accessible spatiotemporal scales occurring in chemical and physical systems.[1-10] This enhancement is feasible by integrating out fast degrees of freedom at the fine-grained (FG) resolution, thus extending the applicability of computer simulations to larger system sizes and longer simulation times than was previously perceived to be possible (for example, from HIV-1 research, see Refs. 11-15). To reproduce equilibrium structural correlations in CG models compared to the FG reference, various bottom-up CG methods have been proposed to approximate the potential of mean force (PMF) between the CG particles as the effective CG interaction.[1-10, 16-20] Since the static equilibrium correlations can be well-captured by designing these conservative interactions between CG particles, bottom-up CG simulations are generally carried out with Newtonian dynamics having only conservative forces. During this CG Newtonian mechanics, the CG configuration space tends to have a flatter or softer effective potential energy surface than that of FG reference, resulting in faster diffusion dynamics than in the FG dynamics.[21-26] While this artificial "acceleration" helps to extend the sampling of the CG simulation, it hinders the study of explicitly dynamical properties with the CG model. It is known from the Mori-Zwanzig formalism[27-30] that this acceleration in the CG model occurs due to missing fluctuation and dissipation forces in the CG equations of motion, but there is no alternative theory to qualitatively understand the physical nature of these missing forces and to quantitatively estimate the degree of the acceleration in the CG model over the FG one due to the absence of fluctuation and dissipation forces.

In this light, numerous approaches and theories have been developed and applied to understand the CG dynamics with respect to the reference FG dynamics. Briefly, using the projection operator technique of Zwanzig,[27-30] one can show that the time evolution of specific collective variables in the reduced phase space follows the form of the Generalized Langevin Equation (GLE). Therefore, in an ideal sense, dynamics in the reduced CG representation could be faithfully represented by the Mori-Zwanzig formalism. Despite its theoretical rigor, a full derivation of Mori-Zwanzig equations of motion for the CG system would be highly challenging or even intractable without making approximations to specify the non-Markovian frictional and fluctuation forces, which in turn are completely dependent on the nature of the system.[23, 31] While we are aware of numerous efforts to emulate both the structural and dynamical properties of CG systems based on the Mori-Zwanzig formalism, e.g., works from Español, Karniadakis, and others,[23, 25, 26, 32-47] we instead focus here on the CG Newtonian mechanics and aspire to better understand and then to characterize and model the acceleration behavior of the CG system over the FG one from which it is derived.

In the literature, several researchers have proposed a time-rescaling approach where the time step in CG simulations is effectively rescaled to match some reference dynamical behavior.[24, 48-50] While time-rescaling offers a reasonably intuitive and sometimes applicable approach, there are intrinsic drawbacks, such that the scaling behavior is not universal among different systems.[51-54] In order to exactly determine the rescaling factor, one may need to revisit the GLE or Langevin Equation to estimate the frictional forces.[35] A comprehensive review of these distinct approaches to understanding the dynamics of CG models is detailed in the literature.[55, 56]

In this series of papers, we offer a different approach to assessing the dynamical properties of the CG system. This alternative approach is based on the excess entropy of the system. The relationship, known as "excess entropy scaling", posits that the dynamical properties of a system (especially self-diffusion) scale as the excess entropy of that system.[57-61] Even though suggesting that entropy as a thermodynamic quantity is related to the system dynamics seems counterintuitive at first glance based on fundamental statistical mechanics,[62] this semi-quantitative relation has been established to be useful under certain systems:[63] Starting from the original work for the Lennard-Jones and soft-sphere models,[57] the excess entropy scaling formalism has been demonstrated in many complex systems such as liquid metals,[64, 65] ionic liquids,[66] polymer chains,[67] and even extended to confined systems.[68, 69] We note that these "liquid" systems are at ambient (or normal liquid, not deeply supercooled, regime) conditions at relatively high temperatures that do not exhibit strong activated dynamics. Given its apparent validity in a number of physical systems, several theoretical developments have provided some physical explanations underlying the excess entropy scaling relationship. As hinted by mode coupling theory (MCT),[70, 71] that links structures and dynamics based on the radial distribution function (RDF),[72, 73] recent work has further suggested that the phenomenological entropy scaling can be microscopically derived in certain conditions.[74] Also, other theoretical connections between the Kolmogorov-Sinai entropy and dynamical properties are described in the literature.[73-77] It should be noted that recent findings suggest that one can possibly derive the scaling relationship in simple Hamiltonian systems using the Boltzmann's formula and rate theory.[61] Yet, there appears to be no systematic, first-principles theory to fully derive the excess entropy scaling relationship in any Hamiltonian system to date.

Interestingly, relatively less attention has been given to applications of excess entropy scaling to CG models, and therefore it is of particular interest to systematically extend this relationship to CG dynamics. While earlier work provided somewhat limited interpretations[78] or implications for the relative entropy formalism of simple models,[79] recent research efforts to apply the excess entropy scaling to CG models have been reported,[80, 81] hinting that a rigorous extension to CG systems may be feasible. However, two main caveats have limited the application of this concept to CG modeling. First and foremost, despite its applicability for a wide range of molecular systems, there is no guarantee that such a scaling relationship still holds for CG systems. More importantly, even if it does hold, no systematic theories or studies have shown that both FG and corresponding CG systems will exhibit the same (universal) scaling behavior. Without confirming the universality between FG and its corresponding CG systems, an effort that tries to understand CG dynamics in terms of excess entropy scaling would be difficult. In particular, the aforementioned drawbacks are due to the empirical nature of the scaling relationship, and thus the physical understanding and correspondence between full FG dynamics and CG dynamics are still premature and largely unexplored.

In this series of papers, we aim to deliver a more comprehensive understanding of CG dynamics with respect to the underlying FG dynamics. For this first paper of the series, we will introduce the concept of excess entropy scaling for the FG system and then also the corresponding CG system by developing a new method to correctly address different modal contributions to excess entropy. Then, we will determine the excess entropy of atomistic systems and their CG counterparts based on entropy representability. After confirming the excess entropy scaling relationship holds in general, we will explore if the same scaling relationship holds for both the FG and CG dynamics for the same molecular system in order to elucidate the universal nature of excess entropy scaling in such systems.

## II. Theory
### A. Excess Entropy Scaling Relationship

The starting point for the scaling relationship is to define the excess entropy of the system at a given number density $\rho$ and temperature $T$ such that

$$S_{ex} = S_{ex}(\rho, T) := S(\rho, T) - S_{id}(\rho, T). \tag{1}$$

Since the ideal gas is the maximally disordered state, this condition gives $S_{ex} < 0$ for non-ideal systems. The excess entropy scaling aims to connect this thermodynamic property to a reduced dynamic quantity such as the self-diffusion coefficient. Two distinct scaling schemes have been suggested, and here both are introduced and compared comprehensively.

First, the most well-known scaling relationship, known as the Rosenfeld scaling, is given by:[57-59]

$$D^* = D_0 \exp(\alpha S_{ex}/Nk_B), \tag{2}$$

where the reduced dynamic property (diffusion coefficient in this case) is scaled as

$$D^* = D \frac{\rho^{\frac{1}{3}}}{\left(\frac{k_B T}{m}\right)^{\frac{1}{2}}}. \tag{3}$$

Here, $\alpha$ and $D_0$ are the coefficients obtained from the scaling relationship and are dependent on the system of interest. Alternatively, Dzugutov reported a similar relationship expressed by a slightly different formalism:[60]

$$D_Z^* = D_Z^0 \exp(S_{ex}^{(2)}/Nk_B), \tag{4}$$

where $S_{ex}^{(2)}$ is the two-body contribution to the full excess entropy $S_{ex}$ including multi-particle correlations. Even though Eq. (2) and (4) look similar at first glance, the two scaling schemes have distinct $D_0$ and $\alpha$ expressions. Such differences between the Rosenfeld and Dzugutov scaling schemes are extensively discussed in the literature, but herein we emphasize the key differences and determine which scheme is more physically appropriate for the scope of this work. First, in Eq. (4), the diffusion coefficient is reduced by $D_Z^* = D \cdot (\sigma^{-2} \Gamma_E^{-1})$ using the Enskog theory for hard spheres with the collision frequency of $\Gamma_E = 4\pi\sigma^2 \rho g(\sigma) \sqrt{k_B T/(\pi m)}$. The idea of using the microscopic Enskog collision rate is because the Dzugutov scaling was derived by assuming that the diffusion process can be regarded as a molecular "caging" effect in hard spheres.[60] In this sense, the exponent in the Dzugutov scaling becomes unity ($\alpha = 1$) under the ergodic assumption, but this scaling may only be valid in hard sphere diffusion processes. We note that recent work demonstrated that this ergodic assumption is usually not satisfied in many Molecular Dynamics (MD) simulations,[82] and an alternative exponent 2/3 was instead derived in the case of the Dzugutov scaling from the MCT.[72] On the other hand, the Rosenfeld scaling applies to many dynamic properties, including diffusion, viscosity, and thermal conductivity, indicating that the Dzugutov scaling is less universal and less globally accurate than the Rosenfeld scaling, as has been pointed out by several researchers.[83]

The physical consistency can also be seen from the scaling coefficient. In Eq. (3), the diffusion coefficient is scaled by elementary quantities (the so-called macroscopically reduced units) that correspond to the characteristic length and timescales in Newtonian dynamics. In theory, the reduced units applied in Eq. (3) are equivalent to the reduced quantities used in the isomorph theory of Dyre, implying a hidden scale invariance of strongly correlating systems, which can explain the quasi-universality of scaling relationships.[84, 85] On the other hand, the reduced unit in the Dzugutov scaling is rather microscopic and empirical, and this scaling scheme may be dubious in some cases where diverging behavior is observed.[77]

Lastly, the excess entropy terms utilized in the two scaling relationships are different. The Dzugutov scaling only utilizes the pair excess entropy term $S_{ex}^{(2)}$ while the other contributions are neglected, whereas the Rosenfeld scaling employs the overall excess entropy of the system $S_{ex}$ to better capture the full entropy contribution.

To summarize, in this series of papers, we follow the original scaling idea from Rosenfeld using the macroscopically reduced units for both FG and CG systems since it is a more general scaling relation for liquids and also consistent with a macroscopic point of view. However, a physical idea behind the Dzugutov scaling, the Enskog picture, will be revisited in the following paper of the series (Paper II). With this in mind, we aim to relate the FG diffusion relationship based on the Rosenfeld excess entropy scaling

$$D_{FG}^* = D_0^{FG} \exp(\alpha^{FG} s_{ex}^{FG}),$$

(5)

to its CG counterpart

$$D_{CG}^* = D_0^{CG} \exp(\alpha^{CG} s_{ex}^{CG}),$$

(6)

where we denote the per particle entropy as $s_{ex} = S_{ex}/Nk_B$ for simplicity. In other words, the main goal of this paper is to determine $\alpha^{FG}$ and $\alpha^{CG}$ by calculating $s_{ex}^{FG}$ and $s_{ex}^{CG}$ to probe whether the CG scaling relationship follows the same scaling exponent. The theories behind the calculation of $s_{ex}^{FG}$ and $s_{ex}^{CG}$ are presented in the next section.

## B. Excess Entropy Estimation
### 1. Conventional Approach: Multiparticle Correlation Expansion

Substantial research effort has focused on the means for calculating excess entropy. In principle, a direct determination of excess entropy is possible by separately calculating the system entropy and the ideal gas entropy using conventional methods, e.g., thermodynamic integration (TI). Although we are aware of many papers that have employed TI to calculate excess entropies,[86-88] we opt not to use this method in this work because TI is less straightforward for linking the thermodynamics and structures between the detailed (FG) system to its reduced (CG) version. We seek to utilize a more straightforward method that can physically elucidate the missing degrees of freedom during the coarse-graining process in order to estimate excess entropy. Additionally, there are several practical concerns of conventional TI calculations: (1) design of a corresponding ideal gas system is needed, and (2) slow numerical convergence of the method.

Instead of the TI method, a configuration-based approach to calculating the excess entropy quantity can be designed based on statistical mechanical theory. From Green's derivation of multiparticle correlation functions,[89] Wallace rederived a systematic expansion of the excess entropy expressed as a sum of integrals of *n*-particle distribution functions:[90-93]

$$S_{ex} = \sum_{n \geq 2} S^{(n)}.$$

(7)

In Eq. (7), $S^{(n)}$ is the excess entropy contribution from the *n*-particle contributions. The simplest and the most common approach to evaluating $S_{ex}$ is to approximate the two-body contribution as being dominant compared to other higher-order contributions, as shown in several theoretical derivations.[89, 94-96] Furthermore, for a variety of systems, this assumption has been widely adopted due to the fact that the $S^{(3)}$ term cancels other higher-order contributions.[66, 97-100] With this in mind, the $S^{(2)}$ term is written as a function from the pair distribution function $g^{(2)}(\mathbf{r})$

$$S^{(2)} = -2\pi \int_0^\infty \{g^{(2)}(\mathbf{r}) \ln g^{(2)}(\mathbf{r}) - [g^{(2)}(\mathbf{r}) - 1]\} \mathbf{r}^2 \cdot d\mathbf{r},$$

(8)

where the vector **r** includes information about both the position and orientation of the molecule. Namely, Equation (8) provides a configurational basis to estimate excess entropy for various systems of interest.

## 2. Orientational Contribution

An important point to note in Eq. (8) is that the pair distribution function $g^{(2)}(\mathbf{r})$ is a function of not only the pair distance between particles $r = |\mathbf{r}|$ but also positions of the particle pair $\mathbf{r}_1$ and $\mathbf{r}_2$ with their orientations $\omega_1$ and $\omega_2$, resulting in twelve variables. However, the majority of current literature only considers the translational component while ignoring the orientational (or angular) contribution when calculating the scalar term:[101]

$$S^{(2)}_{trans} = -2\pi\rho \int_0^\infty \{g^{(2)}(r) \ln g^{(2)}(r) - [g^{(2)}(r) - 1]\} r^2 \cdot dr.$$

(9)

It is straightforward to see that Eq. (9) corresponds to only the translational component because the orientational dependence on $\omega_1$ or $\omega_2$ no longer appears. Since only the RDF is required, this choice is more numerically feasible and often preferred. Furthermore, the so-called "homogeneity assumption"[67] whereby $S_2$ contributes nearly 80% to the overall excess entropy is satisfied by simple models where there are no orientational degrees of freedom, e.g., Lennard-Jones fluids.[102, 103] However, as Malvaldi and Chiappe have noted, this approximation is often violated when the molecule has non-spherical symmetry.[66] Thus, in the case of water and methanol studied in the present work, such considerations need to be carefully addressed. In particular, many studies reported calculating the excess entropy of water to understand its connection to anomalous properties.[86, 104-107] Subsequently, two pertinent findings were discovered: (1) the contribution of non-translational components in water is large,[108-112] (2) the orientational entropy is not explicitly connected to the translational entropy,[113] meaning that translational entropy based on the RDF cannot account for orientational contributions. Altogether, we believe that having only translational components from the RDF to calculate the excess entropy can incur major errors.

In that light, several theories and computational techniques have been devised to account for the orientational contributions beyond the pair translational motion. For rigid molecules, Lazaridis and Karplus suggested that the function $g^{(2)}(\mathbf{r}_{12})$ can be factorized as follows:[109]

$$g^{(2)}(\mathbf{r}_{12}) = g^{(2)}(r, \omega_1, \omega_2) = g^{(2)}_{\text{trans}}(r) \cdot g^{(2)}_{\text{or}}(\omega_1, \omega_2 | r).$$
(10)

Later, Zielkiewicz[108] designed a slightly different factorization scheme in terms of translational, configurational, and orientational contributions in order to perform feasible integrations:

$$g^{(2)}(\mathbf{r}_{12}) = g^{(2)}(r, \omega_1, \omega_2) = g^{(2)}_{\text{trans}}(r) \cdot g^{(2)}_{\text{conf}}(\omega_1 | r) \cdot g^{(2)}_{\text{orient}}(\omega_2 | r, \omega_1).$$
(11)

Equations (10)-(11) separate the translational component from the orientational contribution by using a conditional probability distribution function. However, a direct calculation of configurational and orientational correlations of molecular pairs in liquids is numerically challenging due to the multi-dimensional numerical integrations. To elaborate, this orientational contribution is characterized by the following integration form

$$S^{(2)}_{\text{or}} = -2\pi\rho \int_0^\infty g^{(2)}(r) \cdot \left( -\frac{1}{\Omega^2} \int\int J(\omega_1, \omega_2) g(\omega_1, \omega_2 | r) \ln g(\omega_1, \omega_2 | r) d\omega_1 d\omega_2 \right) dr,$$
(12)

where $J(\omega_1, \omega_2)$ is the Jacobian of the angular variables $\omega_1, \omega_2$. Under homogeneous and isotropic conditions, we can reduce the angular variables to five[109] or six angles[108] depending on the choice of factorization. However, a full calculation of such collective integrations over both distance and orientational variables is numerically very demanding.[78, 108, 114, 115] Even though several algorithmic improvements have been proposed after Lazaridis, Karplus, and Zielkiewicz,[116, 117] only a few follow-up reports are available in the literature to date. Yet, these approaches emphasize that one must consider more than just the translational contribution, especially in water.

## 3. Alternative Approach: 2PT-Based Method

To surmount the numerical challenges mentioned earlier, we use the two-phase thermodynamic (2PT) method[118-122] that has been applied to various chemical systems[123-127] as an alternative approach. Herein, we provide essential steps in performing the 2PT simulation, with more detailed derivations and discussion can be found in Refs. 118-122. In a nutshell, the 2PT method constructs the partition function of the system by designing the density of the states (DoS) of the system as a linear combination of solid-like and gas-like (diffusive) components. The total DoS of liquid is usually obtained from the Fourier transform of the velocity autocorrelation function (VACF) $C_{vv}(t)$

$$\text{DoS}_{\text{liq}}(v) = \lim_{\tau \to \infty} \frac{1}{2k_B T} \int_{-\tau}^{\tau} C_{vv}(t) e^{-2\pi v t} dt,$$
(13)

where $v$ denotes the frequency of normal modes. This allows for decomposing the liquid DoS into a fully harmonic solid-like component and an anharmonic diffusive component with weights $\omega(v)$

$$\text{DoS}_{\text{liq}}(v) = \omega_{\text{solid}}(v) \text{DoS}_{\text{solid}}(v) + \omega_{\text{gas}}(v) \text{DoS}_{\text{gas}}(v).$$
(14)

Detailed formulas for $\omega_{\text{solid}}(v)$ and $\omega_{\text{gas}}(v)$ can be found in the original 2PT literature.[118-122] From the decomposed liquid DoS in Eq. (14), the corresponding thermodynamic entropy can be calculated using the entropy weighing function $W(v)$ for solid and gas phases. For example, the gas component as a hard sphere system such that the gas-like weights $W_{\text{gas}}(v)$ are readily given as a function of the hard sphere entropy. The $\text{DoS}_{\text{gas}}(v)$ term is then calculated by solving the fluidity term in the gas-like DoS by matching the diffusivity of the system to the Carnahan-Starling equation of state. Lastly, the solid-like weight $W_{\text{solid}}(\nu)$ is assumed to be a harmonic oscillator: $W_{\text{solid}}(\nu) = \frac{\beta\hbar\nu}{\exp(\beta\hbar\nu)-1} - \ln[1 - \exp(-\beta\hbar\nu)]$.

This basic approach of 2PT is then applied to the primary degrees of freedom in the molecular frame in order to decompose thermodynamic properties into different modal contributions. In other words, the overall velocity of molecules can be decomposed in the same manner
$$v(t) = v_{\text{trans}}(t) + v_{\text{rot}}(t) + v_{\text{vib}}(t).$$
(15)

While the translational velocity corresponds to the center-of-mass motion, the rotational velocity can be determined from the angular velocity $\omega(t)$ based on the definition of the angular momentum $L = \mathbf{I}\omega$. By inverting the inertia tensor $\mathbf{I}$, one could estimate $\omega(t)$, and then the angular velocity $v_{\text{rot}}(t)$ is computed as $v_{\text{rot}}(t) = \omega(t) \times r(t)$. Having these two components in hand, the vibrational velocity $v_{\text{vib}}(t)$ is readily obtained as a complement: $v_{\text{vib}}(t) = v(t) - (v_{\text{trans}}(t) + v_{\text{rot}}(t))$. Then, the thermodynamic quantities corresponding to each motion are calculated based on Eq. (14). Assuming this general decomposition of atomistic velocities, the overall entropy of the system can be decomposed into translational, rotational, and vibrational components as[118-127]
$$S = S_{\text{trans}} + S_{\text{rot}} + S_{\text{vib}}.$$
(16)

Each entropy component ($t \cup r \cup v$ indicating each modal contribution, respectively) is calculated by
$$S_{t \cup r \cup v} = k_B \left[ \int_0^\infty W_{\text{solid}}^{t \cup r \cup v}(\nu) \text{DoS}_{\text{solid}}^{t \cup r \cup v}(\nu) + \int_0^\infty W_{\text{gas}}^{t \cup r \cup v}(\nu) \text{DoS}_{\text{gas}}^{t \cup r \cup v}(\nu) d\nu \right],$$
(17)

meaning that one must determine $W_{\text{solid}}^{t \cup r \cup v}(\nu)$, $\text{DoS}_{\text{solid}}^{t \cup r \cup v}(\nu)$, $W_{\text{gas}}^{t \cup r \cup v}(\nu)$, and $\text{DoS}_{\text{gas}}^{t \cup r \cup v}(\nu)$. First, translational entropy is calculated by using the center-of-mass velocities to construct the $\text{DoS}_{\text{liq}}^{\text{COM}}(\nu)$ and to apply the aforementioned procedures to yield $W_{\text{solid}}^{\text{COM}}(\nu)$ and $W_{\text{gas}}^{\text{COM}}(\nu)$. For the rotational components, we repeat the calculation using $v_{\text{rot}}(t) = \omega(t) \times r(t)$. Finally, the vibrational entropy is computed solely from the solid-like contributions of the harmonic oscillators.

The partition scheme that 2PT provides as translational, rotational, and vibrational contributions is also consistent with the partition scheme used in Eqs. (10)-(11). We envisage that the translational contribution remains the same, while the rotational and vibrational components complement the orientational entropies. Moreover, the 2PT partition scheme, Eq. (16), allows the differentiation of the missing modes from the CG procedure. It is worth noting that we recently elucidated the missing entropies during the coarse-graining process for the single-site CG model using this framework.[128] Here, single-site CG models are represented as the center-of-mass of each molecule, which interact via isotropic CG pair potentials. At this CG resolution, assessing the missing entropy is much more intuitive since the CG system has only translational motions. The entropy representability relationship further suggests that the translational components from the FG entropy remain at the CG resolution, while other entropies from rotation and vibration are mapped into the approximate CG PMF.[128] A theoretical link between the missing entropy and excess entropy scaling was first suggested by Ref. 79 using an idea of relative entropy, where it was postulated that the relative entropy differences are responsible for differences in dynamics if the scaling relationship in both models holds with the same scaling exponent. However, in their work,[79] this hypothesis was originally suggested for analytically mapping Lennard-Jones liquids to simple soft-sphere potentials at the same resolution and was never rigorously verified for actual molecular systems upon a coarse-graining process.

Despite the 2PT receiving less attention in literature for the calculation of excess entropy, a few examples have successfully demonstrated its applicability. To our knowledge, employing the 2PT-based approach for the atomistic liquid systems was first introduced by Dhabal et al.,[107] and then Palomar and Sesé separately demonstrated the 2PT-based approach for atomistic dipolar liquids.[129] More importantly, these two distinct efforts were able to show that the 2PT-based excess entropy reproduces trends as accurately as from the TI values, thus validating its feasibility in computing excess entropy. Similarly, work from Marvin et al. utilized the 2PT method as an extension of Ref. 128 to obtain the translational excess entropies in a CG model.[81] Both papers illustrated that the 2PT method could be used to calculate the excess entropy of the CG system; however, the aforementioned attempts were only limited to translational motion in the center-of-mass frame. We note that considering only a translational component to the excess entropy ignores contributions from other motions at resolutions finer than the single-site CG, resulting in an incorrect scaling relationship and dynamics. The $\alpha$ exponent near 0.38 in Eq. (2) for CG systems was underestimated as reported in Ref. 81. This value suggests that one needs to fully consider the representative motions at the CG level. Inspired by this inconsistency, we propose a full 2PT-based approach to calculating the excess entropy beyond the translational entropy for both the FG and its corresponding CG systems by addressing entropy representability. While one could recover the missing entropy values from CG simulations by utilizing entropy representability,[128] here our intention is to elucidate the excess entropy differences between the FG and CG systems since most CG simulations with accelerated CG dynamics do not fully account for the missing entropy.

Our proposed approach utilizes the same partitioning used in the 2PT method. That is, we divide the FG system entropy into translational and angular terms

$$S^{\text{FG}} = S^{\text{FG}}_{\text{trn}} + S^{\text{FG}}_{\text{or}} = S^{\text{FG}}_{\text{trn}} + S^{\text{FG}}_{\text{rot}} + S^{\text{FG}}_{\text{vib}}.$$
(18)

The first separation in Eq. (18) was from Eq. (10), and thus the excess entropy at the FG model becomes

$$S^{\text{FG}}_{ex} = S^{\text{FG}} - S^{(\text{id})} = S^{\text{FG}}_{\text{trn}} + S^{\text{FG}}_{\text{rot}} + S^{\text{FG}}_{\text{vib}} - \left(S^{(\text{id})}_{\text{trn}} + S^{(\text{id})}_{\text{rot}} + S^{(\text{id})}_{\text{vib}}\right).$$
(19)

Likewise, the entropy of the corresponding single-site CG system can be formulated as

$$S^{\text{CG}} = S^{\text{CG}}_{\text{trn}},$$
(20)

where the corresponding excess entropy is solely the translational contributions from Eq. (19), i.e.,

$$S^{\text{CG}}_{ex} = S^{\text{CG}} - S^{(\text{id})} = S^{\text{CG}}_{\text{trn}} - S^{(\text{id})}_{\text{trn}}.$$
(21)

To reiterate, the prior work in the literature has mainly attempted to employ Eq. (20) to determine excess entropy scaling, but here we emphasize that one must also consider $S^{\text{FG}}_{\text{rot}} + S^{\text{FG}}_{\text{vib}}$ to correctly address the question of FG scaling. We will demonstrate this method for two different molecular systems in the next section.

**C. Excess Entropy Formulation**
**1. Fine-Grained System**

From the definition of the excess entropy, the FG entropy can be obtained from a direct 2PT calculation, whereas the corresponding ideal gas entropy term can be written as an analytical formulation deduced from statistical mechanical theory.[130] In spite of the FG resolution, the diffusion behavior is characterized by the trajectories of the mapped center-of-mass of each molecule. Therefore, the translational entropy must correspond to the single-component ideal gas, and this term can be readily obtained by employing the Sackur-Tetrode equation

$$s_{trn}^{(id)} = \frac{S_{trn}^{(id)}}{Nk_B} = -\ln\left(\frac{h^2}{2\pi m k_B T}\right)^{\frac{3}{2}} - \ln\left(\frac{N}{V}\right) + \frac{5}{2}.$$

(22)

Even though the translational motions can be extracted from the center-of-mass entities, the rotational motions also contribute to the overall entropy of the system at the FG resolution. For an ideal gas, the rotational partition function for the specific rotational modes at a temperature $\Theta_r$ has the form

$$q_{rot} = \frac{8\pi^2 I k_B T}{\sigma h^2} = \frac{T}{\sigma \Theta_r}.$$

(23)

The characteristic rotational temperatures are related to the moment of inertia with respect to the principal moments $I_r$: $\Theta_r = h^2/8\pi^2 k I_r$. From the rotational partition function, the corresponding ideal gas entropy term is given as

$$s_{rot}^{(id)} = \frac{S_{rot}^{(id)}}{Nk_B} = \ln\left[\frac{\sqrt{\pi}}{\sigma}\left(\frac{T^3 e^3}{\Theta_A \Theta_B \Theta_C}\right)^{\frac{1}{2}}\right],$$

(24)

where the rotations along the *x, y,* and *z*-axes have rotational temperatures ordered as $\Theta_A \geq \Theta_B \geq \Theta_C$.

The last term in Eq. (19) that corresponds to the vibrational contributions is computed from the polyatomic vibrational partition function

$$q_{vib} = \prod_j^{F_v} \frac{e^{-\Theta_{v_j}/2T}}{1 - e^{-\Theta_{v_j}/T}},$$

(25)

where the characteristic vibration temperatures are similarly defined as $\Theta_{v_j} = h\nu_j/k$ by denoting $\nu_j$ as the characteristic frequency of the *j*th vibrational modes. $F_v$ are the remaining degrees of freedom for the vibrational modes. For our work, $F_v = 3N - 6$ due to non-linearity. Altogether, $q_{vib}$ determines the vibrational entropy of the ideal gas $s_{vib}^{(id)}$

$$s_{vib}^{(id)} = \frac{S_{vib}^{(id)}}{Nk_B} = \sum_{i=1}^{3N-6}\left[\frac{\frac{\Theta_{vj}}{T}}{e^{\frac{\Theta_{vj}}{T}} - 1} - \ln\left(1 - e^{-\Theta_{vj}/T}\right)\right].$$

(26)

However, vibrational contributions are generally known to be negligible compared to translational and rotational components. Especially, for FG water force fields, the oxygen-hydrogen bond is usually considered rigid. Therefore, the final excess entropy term used in this work is written as

$$s_{ex}^{FG} = s^{FG} - \left[\frac{5}{2} - \ln\left(\frac{h^2}{2\pi m k_B T}\right)^{\frac{3}{2}} - \ln\left(\frac{N}{V}\right)\right] - \ln\left[\frac{\sqrt{\pi}}{\sigma}\left(\frac{T^3 e^3}{\Theta_A \Theta_B \Theta_C}\right)^{\frac{1}{2}}\right].$$

(27)

### 2. Coarse-Grained System

Equation (21) is built upon the findings of Ref. 128 that the single-site CG model has only translational motions. Since the CG system is constructed under the same condition as the FG system, we can repeat Eq. (22) for the CG excess entropy $S_{ex}^{*CG}$ as

$$s_{ex}^{CG} = s^{CG} - \left[\frac{5}{2} - \ln\left(\frac{h^2}{2\pi m k_B T}\right)^{\frac{3}{2}} - \ln\left(\frac{N}{V}\right)\right].$$

(28)

The CG excess entropy becomes fairly complicated when each molecule is mapped to more than two CG sites. This mapping is beyond the scope of the current paper, but here we note that one should include the rotational contribution to the CG excess entropy by extracting the angular velocity of the CG beads along with the computed moment of inertia tensor following the 2PT argument. A detailed investigation of these cases is underway.

### D. Coarse-Grained Model
### 1. CG Water Model: BUMPer

An accurate bottom-up CG model is designed to capture important structural correlations in the CG system.[8, 131-133] Among a number of bottom-up CG methods, the Multiscale Coarse-Graining (MS-CG) methodology is shown to capture up to three-body correlations using only two-body basis sets. Practically speaking, the MS-CG force fields are fitted to approximate the many-body PMF using variational force-matching. In this regard, CG water models often aim to capture three-body correlations in water that originate from hydrogen bonding.[134-136] However, in the force-matched CG water models, we recently found that the pairwise basis sets fail to reproduce structural correlations, unlike other simple liquids. An alternative approach is to utilize the CG water model based on the Stillinger-Weber (SW) interaction,[137] such as the mW model,[138] or the three-body force-matched CG model.[139, 140] However, having a three-body interaction would slow down the overall performance of the CG model, and there have been studies indicating that the SW-based water model may be problematic for the entropy scaling due to its many-body nature.[64, 141, 142]

The aforementioned drawbacks can be mitigated by employing our recently developed bottom-up CG model, which we call the Bottom-Up Many-Body Water (BUMPer) model.[143, 144] We have shown that the BUMPer CG interaction faithfully corresponds to the CG interaction obtained from many-body projection theory. In brief, the many-body projection theory stems from the many-body expansion (MBE) of the CG interaction $U(\mathbf{R}^N)$

$$U(\mathbf{R}^N) = \sum_I \sum_{J \neq I} U^{(2)}(R_{IJ}) + \sum_I \sum_{J \neq I} \sum_{K > J} U^{(3)}(\theta_{JIK}, R_{IJ}, R_{IK}) + \sum_{IJKL} U^{(4)}(\theta_1, \theta_2, \phi) + \cdots,$$

(29)

to effectively project higher-order interactions onto lower-order basis sets, which is similar to the Bogoliubov-Born-Green-Kirkwood-Yvon hierarchy in liquid state theory.[145-149] We applied this theory by first performing force-matching up to three-body interactions,[140] and then effectively projecting the force-matched three-body SW interaction $U^{(3)}(\theta_{JIK}, R_{IJ}, R_{IK})$, which is written as

$$U^{(3)}(\theta_{JIK}, R_{IJ}, R_{IK}) = \lambda_{JIK}(\cos\theta_{JIK} - \cos\theta_0)^2 \exp\left(\frac{\gamma_{IJ}}{R_{IJ} - \sigma_{IJ}}\right) \exp\left(\frac{\gamma_{IK}}{R_{IK} - \sigma_{IK}}\right), \tag{30}$$

onto the pairwise basis sets via

$$U(\mathbf{R}^N) = \sum_I \sum_{J \neq I} \left\{ U_{3b}^{(2)}(R_{IJ}) + 2(N_c - 1) \int d\theta_{JIK} dR_{IK} p(\theta_{JIK}, R_{IK}|R_{IJ}) U^{(3)}(\theta_{JIK}, R_{IJ}, R_{IK}) \right\}. \tag{31}$$

We will illustrate the physical principles of the force-matching methodology in the next subsection. From Eq. (31), we demonstrated that the BUMPer model could faithfully recapitulate the structural correlations of water such as two-, three-, and multi-body correlations while retaining pairwise basis sets at an inexpensive computational cost.[143, 144] Thus, we envisage that the excess entropy scaling relationship will still hold in the CG water system while recapitulating the structural correlations in water. Readers are referred to Refs. 143, 144 for the BUMPer theory and detailed discussion.

## 2. CG Methanol Model

The CG model for methanol is generated by mapping the methanol molecule to the center-of-mass at a single-site level. Then, the CG interaction parameters between the methanol CG sites are parametrized by the Multiscale Coarse-Graining (MS-CG) method.[150-154] In a practical sense, the force-matching technique was employed by matching the mapped FG forces exerted on the center-of-mass $\mathbf{f}_I(\mathbf{R}^N)$ to the CG force $\mathbf{F}_I(\mathbf{R}^N)$. This is variationally determined by minimizing the force residual functional $\chi^2[\mathbf{F}]$ defined as

$$\chi^2[\mathbf{F}] = \frac{1}{3N} \langle \sum_{I=1}^N |\mathbf{f}_I(\mathbf{R}^N) - \mathbf{F}_I(M_\mathbf{R}^N(\mathbf{r}^n))|^2 \rangle, \tag{32}$$

where $\mathbf{r}^n$ and $\mathbf{R}^N$ denote FG and CG configurational variables linked by the mapping operator $M_\mathbf{R}^N : \mathbf{r}^n \to \mathbf{R}^N$. Here, $\mathbf{f}_I(\mathbf{R}^N)$ denotes the microscopic force acting on the set $\mathcal{I}_I$ of FG particles $i$ that are mapped into the CG particle $I$ via $M_\mathbf{R}^N$, i.e., $\mathbf{f}_I(\mathbf{R}^N) = \sum_{i \in \mathcal{I}_I} \mathbf{f}_i(\mathbf{r}^n)$. Specifically, CG methanol force fields are also spanned over pairwise basis sets $\{\phi_2\}$ as in the CG water force fields. Unlike water, it has been shown that the pairwise representation of methanol interactions can sufficiently capture the structural correlations at a renormalized CG resolution.[150, 153, 155] Hence, the many-body projection theory accompanied by BUMPer is not needed for methanol. Instead, we expand the interaction by adopting pairwise decomposition as follows

$$\mathbf{F}_I(M_\mathbf{R}^N(\mathbf{r}^n)) = \mathbf{F}_I(\mathbf{R}^N) = \sum_J \phi_2(R_{IJ}) \cdot \hat{e}_{IJ}. \tag{33}$$

In practice, we utilize B-splines to approximate the pairwise basis sets $\phi_2(R_{IJ}) = \sum_k c_k u_k(R_{IJ})$ with respect to the unit vector $\hat{e}_{IJ}$. Therefore, Eq. (32) imparts the optimized spline coefficients $\{c_k\}$ that minimizes the $\chi^2[\mathbf{F}]$.

## E. Computational Details

All simulations in this work were performed by employing the Large-scale Atomic/Molecular Massively Parallel Simulator (LAMMPS) MD engine.[156-158] The FG and CG systems of water[143, 144] and methanol[128] were prepared based on our previous works. In detail, the water system is composed of 512 molecules, and the methanol system consists of 1000 molecules after the energy-minimization process.

Atomistic force fields were chosen as follows: SPC/E,[159] SPC/Fw,[160] TIP4P/2005,[161] TIP4P/Ice[162] for water and OPLS/AA for methanol.[163, 164] The equilibrium system size is chosen from the FG simulations at 300 K and 1 atm condition. This is done by initially annealing the relaxed system using constant $NVT$ dynamics with a Nosé-Hoover thermostat[165, 166] with $\tau_{NVT} = 0.1$ ps for 0.1 ns, followed by constant $NPT$ dynamics for 1 ns using Andersen barostat.[167] At the target temperature, we generated the FG trajectories for 5 ns under constant $NVT$ dynamics using the same settings.

From the $NVT$ trajectories, the effective CG interactions of BUMPer were parametrized based on Refs. 143, 144. For methanol, Eq. (33) was implemented by employing the 6$^{th}$ order B-splines with a resolution of 0.20 Å. The obtained inner-core interactions were additionally fitted to $A \cdot \mathbf{R}^{-B}$ form to account for a poor sampling in the core area.[168] Using the parametrized CG interactions, we ran the CG simulations for analysis under a constant $NVT$ ensemble with a Nosé-Hoover thermostat[165, 166] for 5 ns. For both FG and CG simulations, we collected the configurations every 1 ps.

In summary, Figure 1 showcases the schematic flowchart of this paper. From the FG (atomistic) simulations, we can calculate the overall entropy and then excess entropy. Applying the Rosenfeld scaling to the excess entropy, we can link this excess entropy to the dynamical information. While the same logic applies to the CG system, in this section we introduced a new method to calculate the excess entropy ② and revisited the entropy representability relationship between the FG and CG counterparts ① and ③. The remainder of this paper will be devoted to examining the suggested theory and methodology.

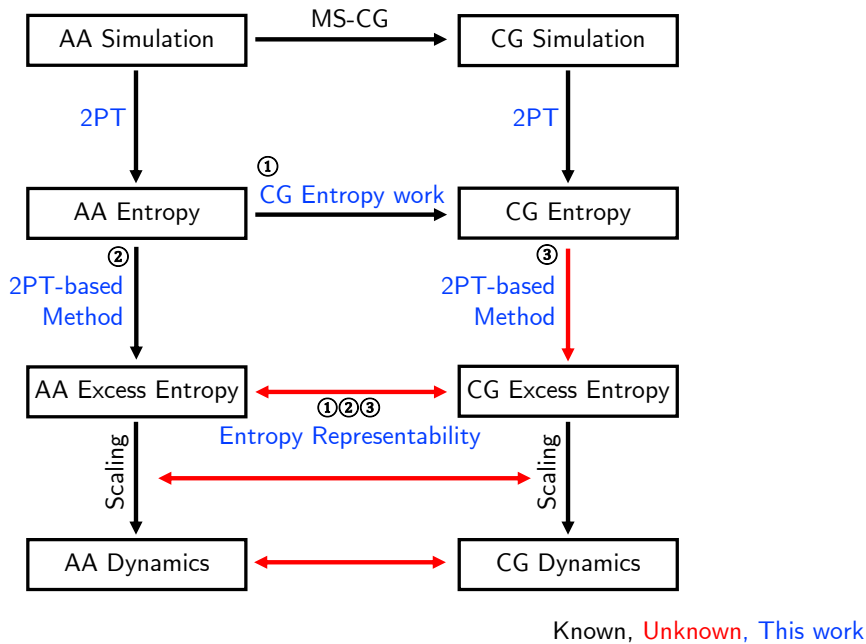

**Figure 1:** Schematic diagram describing the correspondence between the FG and CG dynamics. Dynamical properties are obtained by excess entropy scaling, which is determined by system entropy using the 2PT-based method in this work (blue). Our work aims to understand the unknown links (red) to provide a comprehensive understanding of FG and underlying CG dynamics.

### III. Results and Discussions
#### A. CG Model Interaction

Parametrized CG models are shown in Fig. 2. Since the excess entropy is a function of (number) density and temperature, $S_{ex} = S_{ex}(\rho, T)$, we fixed the volume from the $T = 300$ K and $P = 1$ atm condition while changing temperature to obtain different $S_{ex}$ values. The main reason for fixing the volume is to enforce the constant volume condition for parametrizing the CG interactions. Since bottom-up CG models identify the many-body CG PMF as the CG interaction, the effective CG interaction should also be interpreted as a free energy quantity.[152] In order to consider the changes in CG PMFs at different temperatures, the CG PMFs were designed to be temperature transferable under constant volume, where the pairwise free energy functional can be interpreted as the Helmholtz free energy ($\Delta F = \Delta E - T\Delta S$): $U_{CG}(R) = \Delta E(R) - T\Delta S(R)$,[128, 169, 170] where $\Delta E(R)$ and $\Delta S(R)$ denote the pair energy and entropy functional, respectively. References 10, 128, and 155 discuss this notational choice in detail. Also, due to the energy representability, $\langle \Delta E(R) \rangle_{CG}$ corresponds to the average energy of the FG system, i.e., $\langle U_{FG}(\mathbf{r}^n) \rangle_{FG}$.[171, 172]

Previously, we have demonstrated that the changes in CG PMFs over temperature, characterized as $\Delta U_{CG}(R)/\Delta T$, remain constant across different temperature ranges (280–360 K for water,[143, 144] 250–400 K for methanol[128, 155]), confirming the temperature transferability of CG models. This linear nature of the temperature-dependent nature of CG interactions is plotted in Fig. 2. Moreover, we observe that both liquid CG interactions have double-well interactions. However, in the CG water models, we note a clear separation between two wells with a repulsive barrier near 3.3 Å, suggesting the strongly structured pair correlations compared to liquid methanol. We point the reader to a comprehensive overview and analysis of the BUMPer CG interactions in Refs. 143, 144.

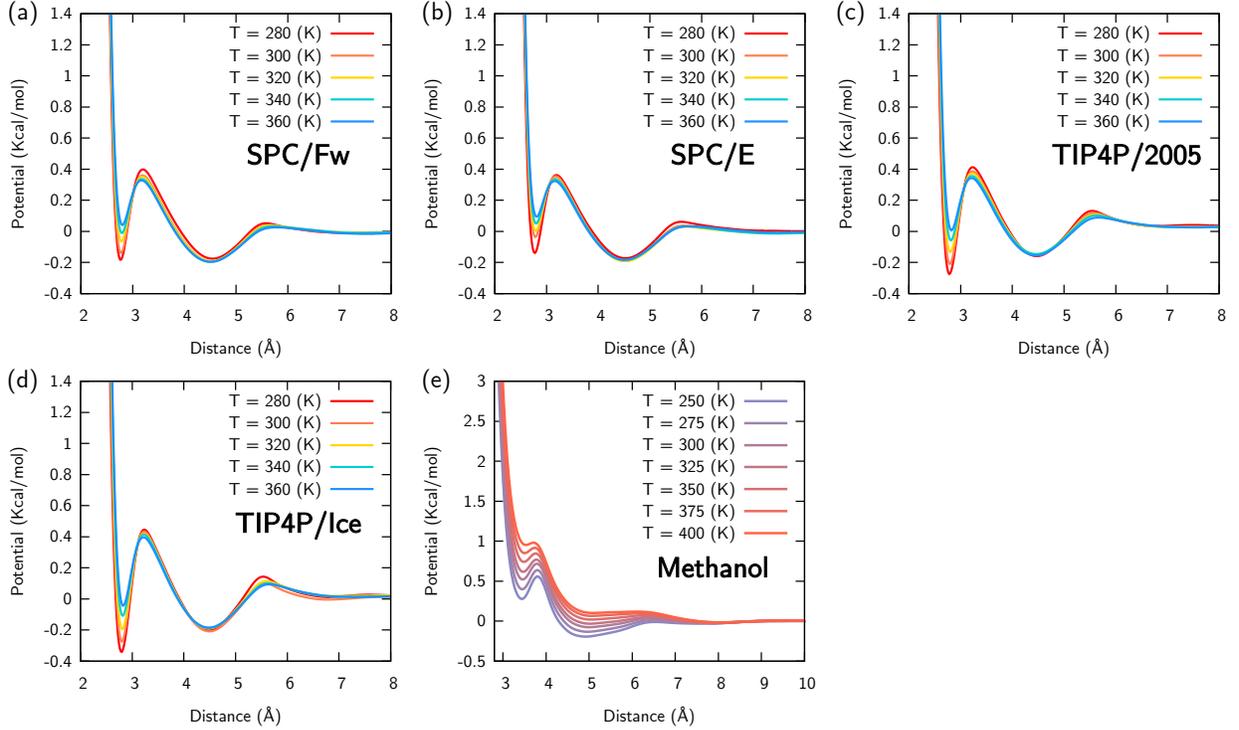

**Figure 2:** Temperature-dependent effective CG pair interactions for the CG model of (a)-(d) water and (e) methanol. For water, different BUMPer models were constructed based on the atomistic force field: (a) SPC/E, (b) SPC/Fw, (c) TIP4P/2005, (d) TIP4P/ice with different temperatures ranging from 280 K (red) to 360 K (blue). CG methanol models are also constructed for different temperatures from 250 K (blue) to 400 K (red).

**B. Diffusion in FG/CG Models**

From the FG simulations and corresponding CG trajectories at different temperatures, we computed the self-diffusion coefficients for water and methanol. The molecular self-diffusion coefficients were readily calculated from the center-of-mass mean squared displacement (MSD)

$$\langle R^2(t)\rangle := \langle |\vec{R}(t) - \vec{R}_0|^2 \rangle = \frac{1}{N}\sum_I^N |\vec{R}_I(t) - \vec{R}_I(0)|^2,$$

(34)

And, thus, from Einstein's relation, we arrive at the diffusion coefficient given as

$$D = \lim_{t\to\infty} \frac{1}{6t}\langle R^2(t)\rangle.$$

(35)

We also alternatively calculated $D$ from Green-Kubo (GK) formalism using the VACF $C_{vv}(t)$:[173]

$$D = \frac{1}{3}\int_0^\infty C_{vv}(t)dt = \frac{1}{3}\int_0^\infty \langle \vec{v}(t)\cdot\vec{v}(0)\rangle \, dt.$$

(36)

Determination of the $C_{vv}(t)$ and the GK calculation were performed in the 2PT program suite, where the velocity autocorrelation was computed by averaging the *x*-, *y*-, and *z*-components of velocity due to isotropy. Even though both approaches give similar values, GK diffusion coefficients are relatively larger and less converged due to the convergence issues of the VACF as reported in the 2PT benchmark paper.[120] Based on this finding, we herein report the diffusion coefficients computed from the MSD.

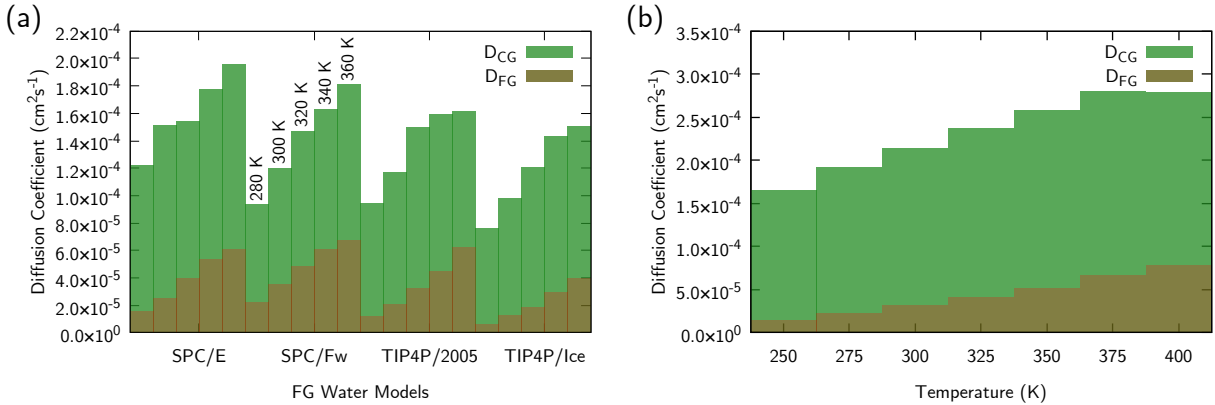

**Figure 3:** Diffusion coefficients of water and methanol evaluated for FG (brown bars) and CG (green bars) systems at different temperatures in each column. (a) Water: SPC/E, SPC/Fw, TIP4P/2005, and TIP4P/Ice force fields from 280 K to 360 K at 20 K intervals. (b) Methanol: OPLS-AA force field at a temperature range from 250 K to 400 K at 25 K intervals.

Figure 3 shows the calculated diffusion coefficients of water and methanol for different FG and CG models at various temperature ranges. We used a range between 280–360 K for water and 250–400 K for methanol. The trend of the diffusion coefficients indicates that the temperature ranges studied here fall into the normal liquid regime. For both molecular systems, we observe that the FG diffusivity increases with temperature as expected, and this trend is also seen in the CG force fields. For water, this trend is invariant under different choices of FG force fields as well. However, we note that the FG diffusion coefficients among the four FG models are slightly different. Even though this mismatch stands out in the case of TIP4P/Ice, this can be understood from the fact that the TIP4P/Ice (unlike the SPC/E model) was designed to match the melting point of the ice, not dynamical properties.[162]

As discussed earlier, the CG dynamics in the Hamiltonian system are generally significantly faster than the FG dynamics, and this acceleration factor is quantified as $D_{CG}/D_{FG}$. Figure 3(a) indicates that $D_{CG}/D_{FG}$ of water monotonically decreases as temperature increases from 280 K to 360 K. Roughly speaking, $D_{CG}/D_{FG}$ falls from a factor of 8–12 at 280 K to 3–4 at 360 K. This trend seems consistent regardless of the FG force fields, indicating an Arrhenius-like behavior of the diffusion coefficient for both water and methanol.

In order to perform excess entropy scaling, we rescaled the diffusion coefficient using the elementary or macroscopic units shown in Eq. (3). Since the CG models and dynamical properties are obtained under the constant volume condition, the size of the simulation box was fixed as the $L_{H_2O} = 25.022$ Å for water according to the original paper with $N = 512$.[143, 144] With these conditions, we arrive at the reduced diffusion coefficient of water as a function of temperature as

$$D^* = D \frac{\rho^{\frac{1}{3}}}{\left(\frac{k_B T}{m}\right)^{\frac{1}{2}}} = 1.4880 \times 10^4 \times \frac{D}{\sqrt{T}}.$$

(37)

For methanol, the reduced diffusion coefficient was similarly scaled by the following equation with $N = 1000$ and $L_{\text{MeOH}} = 41.30$ Å

$$D^* = D \frac{\rho^{\frac{1}{3}}}{\left(\frac{k_B T}{m}\right)^{\frac{1}{2}}} = 1.5029 \times 10^4 \times \frac{D}{\sqrt{T}}.$$

(38)

Table 1 lists the reduced diffusion coefficients for both FG and CG models at different temperatures.

**Table 1:** Macroscopically reduced diffusion coefficients of the water and methanol systems at the FG ($D^*_{\text{FG}}$) and CG ($D^*_{\text{CG}}$) resolutions. For water, we scanned the temperature from 280 K to 360 K using four FG force fields and corresponding BUMPer CG models: (a) SPC/E, (b) SPC/Fw, (c) TIP4P/2005, and (d) TIP4P/ice. For methanol (e), we used a temperature ranging from 250 K to 400 K using OPLS-AA force fields and the corresponding MS-CG model.

| (a) SPC/E: Water | | | (b) SPC/Fw: Water | | |
|---|---|---|---|---|---|
| Temperature | $D^*_{\text{FG}}$ | $D^*_{\text{CG}}$ | Temperature | $D^*_{\text{FG}}$ | $D^*_{\text{CG}}$ |
| 280 K | 1.379×10$^{-2}$ | 1.087×10$^{-1}$ | 280 K | 1.955×10$^{-2}$ | 8.349×10$^{-2}$ |
| 300 K | 2.152×10$^{-2}$ | 1.298×10$^{-1}$ | 300 K | 3.031×10$^{-2}$ | 1.030×10$^{-1}$ |
| 320 K | 3.310×10$^{-2}$ | 1.280×10$^{-1}$ | 320 K | 4.022×10$^{-2}$ | 1.225×10$^{-1}$ |
| 340 K | 4.305×10$^{-2}$ | 1.434×10$^{-1}$ | 340 K | 4.910×10$^{-2}$ | 1.314×10$^{-1}$ |
| 360 K | 4.797×10$^{-2}$ | 1.537×10$^{-1}$ | 360 K | 5.312×10$^{-2}$ | 1.419×10$^{-1}$ |
| (c) TIP4P/2005: Water | | | (d) TIP4P/Ice: Water | | |
| Temperature | $D^*_{\text{FG}}$ | $D^*_{\text{CG}}$ | Temperature | $D^*_{\text{FG}}$ | $D^*_{\text{CG}}$ |
| 280 K | 1.070×10$^{-2}$ | 8.368×10$^{-2}$ | 280 K | 5.370×10$^{-3}$ | 6.757×10$^{-2}$ |
| 300 K | 1.812×10$^{-2}$ | 1.003×10$^{-1}$ | 300 K | 1.068×10$^{-2}$ | 8.411×10$^{-2}$ |
| 320 K | 2.705×10$^{-2}$ | 1.247×10$^{-1}$ | 320 K | 1.567×10$^{-2}$ | 1.005×10$^{-1}$ |
| 340 K | 3.590×10$^{-2}$ | 1.284×10$^{-1}$ | 340 K | 2.375×10$^{-2}$ | 1.156×10$^{-1}$ |
| 360 K | 4.864×10$^{-2}$ | 1.268×10$^{-1}$ | 360 K | 3.134×10$^{-2}$ | 1.183×10$^{-1}$ |
| (e) OPLS-AA: Methanol | | | | | |
| Temperature | $D^*_{\text{FG}}$ | $D^*_{\text{CG}}$ | Temperature | $D^*_{\text{FG}}$ | $D^*_{\text{CG}}$ |
| 250 K | 1.354×10$^{-2}$ | 1.566×10$^{-1}$ | 350 K | 4.131×10$^{-2}$ | 2.073×10$^{-1}$ |
| 275 K | 2.005×10$^{-2}$ | 1.736×10$^{-1}$ | 375 K | 5.207×10$^{-2}$ | 2.174×10$^{-1}$ |
| 300 K | 2.743×10$^{-2}$ | 1.859×10$^{-1}$ | 400 K | 5.924×10$^{-2}$ | 2.095×10$^{-1}$ |
| 325 K | 3.427×10$^{-2}$ | 1.980×10$^{-1}$ | | | |

## C. Excess Entropy Scaling: Water

We next utilized the theory described here to calculate the excess entropy and then examine the validity of the Rosenfeld scaling for various state points of water. First, the translational entropy of the FG system in an ideal gas description was calculated using the following constants $N = 512$, $m_w = 18.015 \times 10^{-3}$ kg, and $L_{H_2O} = 25.022$ Å, giving

$$s_{trn}^{(id)} = -\ln\left[\frac{(6.6262 \times 10^{-34} m^2 \cdot kg \cdot s^{-1})^2}{2\pi\left(\frac{18.015 \times 10^{-3} kg}{6.02214 \times 10^{23}}\right) \times 1.381 \times 10^{-23} J \cdot K^{-1} \times T}\right]^{\frac{3}{2}} - \ln\left[\frac{512}{(25.022 \times 10^{-10} m)^3}\right] + \frac{5}{2} = 1.6787 + \frac{3}{2}\ln T.$$

(39)

As expected, the final result of Eq. (39) shows that the ideal gas entropy is only a function of temperature and number density. Next, we utilized Eq. (24) to calculate the rotational excess entropy of water. From the characteristic rotational temperatures $\Theta_A = 40.1$, $\Theta_B = 20.9$, $\Theta_C = 13.4$ K and the $C_{2v}$ symmetry ($\sigma = 2$), we arrive at the following expression

$$s_{rot}^{(id)} = -3.2840 + \frac{3}{2}\ln T.$$

(40)

Combining Eq. (39) and (40), the final analytical expression is written as

$$s_{ex}^{FG} = S^{FG} - \left(-3.2840 + \frac{3}{2}\ln T\right) - \left(1.6787 + \frac{3}{2}\ln T\right).$$

(41)

Figure 4(a) plots $s_{ex}^{FG}$ over $\ln D_{FG}^*$ for different FG force fields. Before examining the scaling behavior, we first verified that the 2PT method provides physically reasonable excess entropy values. In other words, for all the data points shown in Fig. 4(a), we confirmed that $s_{trn}^{ex}|_{FG} = s_{trn}^{FG} - s_{trn}^{(id)} < 0$ and $s_{rot}^{ex}|_{FG} = s_{rot}^{FG} - s_{rot}^{(id)} < 0$, suggesting a well-defined excess entropy.

It is immediately evident that the linear scaling relationship is satisfied in water regardless of the force field or temperature. In particular, this scaling relationship is given by

$$\ln D_{FG}^* = 0.73 \times s_{ex}^{FG} + 2.15.$$

(42)

We note that this exponent is slightly different from the original exponent from Rosenfeld because the original Rosenfeld scaling was computed for the translational component of the system, lacking configurational effects. We believe this deviation is due to the contribution from the local configurational ordering of water, especially rotational motions.

In order to verify the universality of the scaling relationship between FG and CG systems, we compared $s_{ex}^{CG}$ over $\ln D_{CG}^*$ in Fig. 4(b). This is done by calculating the CG excess entropy from translational contributions

$$s_{trn}^{(id)}|_{CG} = 1.6787 + \frac{3}{2}\ln T.$$

(43)

As expected, we find that $s_{\text{trn}}^{ex}|_{\text{CG}} = s_{\text{trn}}^{\text{CG}} - s_{\text{trn}}^{(\text{id})} < 0$ in the CG system as well. Surprisingly, we still see a linear dependency given by the following relation

$$\ln D_{CG}^* = 0.7 \times s_{ex}^{\text{CG}} - 0.35.$$
(44)

The exponent $\alpha$ obtained from the FG and CG seems slightly different at first glance. Our finding here shows different results with respect to the previous work for the diffusion of the FG and CG water systems.[78] In this previous effort, the authors used the Dzgutov scaling [Eq. (4)] instead of the Rosenfeld scaling [Eq. (2)] to obtain a dimensionless diffusion coefficient. As discussed earlier, Eq. (4) was primarily derived for hard sphere diffusion processes only using pairwise entropy information, which does not hold for water due to hydrogen bonding and large contribution beyond pairwise contributions. Also, while Ref. 78 aimed to compute the orientational entropy based on Refs. 108-112, the computed orientational entropy by sampling various angular configurations may pose a convergence issue.[116, 117] We believe these factors in Ref. 78 result in different scaling exponents for FG and CG water, $\alpha^{\text{FG}} = 0.594$ and $\alpha^{\text{CG}} = 2.373$, which significantly deviate from the conventional Rosenfeld scaling relationship. More importantly, an overestimated CG exponent value of about 8 times indicates that the accurate estimation of excess entropy with a correct scaling scheme is necessary for CG systems. By including all contributions to the entropy with a more accurate scaling relationship as done here, we instead conclude that both FG and CG systems have almost identical scaling relationships, where both exponents also correspond to the range of conventional Rosenfeld scaling. This trend is pronounced in Fig. 4(c), where both FG and CG data points follow nearly identical slopes.

At this point, a natural question arises in terms of differences in $\ln D^*$. As envisaged, the CG system has a larger excess entropy (closer to zero) because the CG system does not have rotational excess entropy. Note that excess entropy is always less than zero, and thus having a less negative value contributes to near positive values (the differences between these entropies will be discussed in a later section). Indeed, it would be of great interest if this agreement between $\alpha^{\text{FG}} = \alpha^{\text{CG}}$ still holds in other molecular systems to confirm a general correspondence.

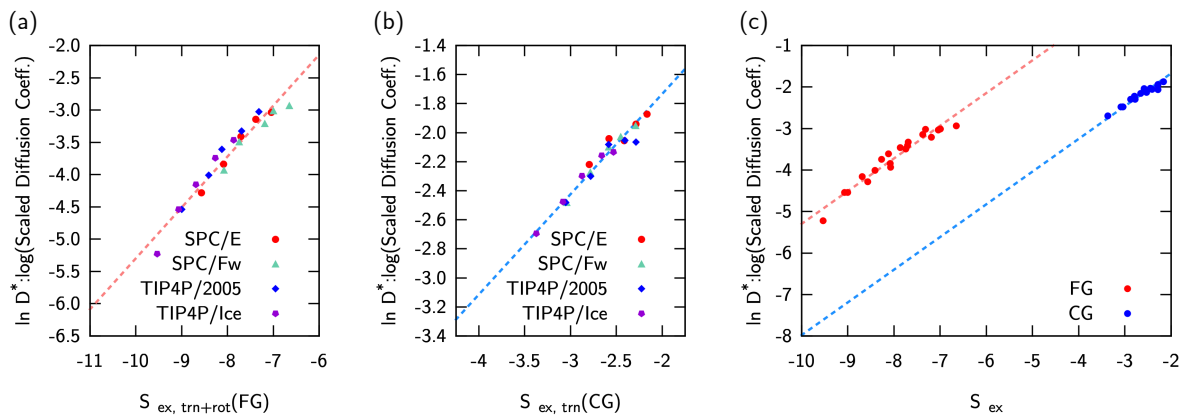

**Figure 4:** Examination of the Rosenfeld scaling in water. (a) FG scaling relationship where the SPC/E (red circle), SPC/Fw (green triangle), TIP4P/2005 (blue diamond), TIP4P/Ice (purple pentagon) models fall into a linear scaling relation shown in Eq. (42) (dashed red line). (b) CG scaling relationship using the BUMPer CG model by parametrizing different force fields. Similarly, a linear excess entropy relationship is obtained (dashed blue line) from Eq. (43). (c) Overall comparison between the FG (red circle) and CG (blue circle) water models. Note that the slope

(exponent) of the FG scaling (dashed red line) is identical to that of the CG scaling relation (dashed blue line), whilst only the *y*-intercept is different.

**D. Methanol**

Given the same scaling behavior obtained in the FG and CG water systems, we now examine the excess entropy scaling of methanol. For the atomistic (FG) methanol, both translational and rotational degrees of freedom contribute to the entropy (vibrational modes are negligible). Using the Sackur-Tetrode equation, the translational entropy of the ideal gas corresponding to the center-of-mass is given as

$$s_{\text{trn}}^{(\text{id})} = -\ln\left[\frac{(6.6262 \times 10^{-34}\text{m}^2 \cdot \text{kg} \cdot \text{s}^{-1})^2}{2\pi \left(\frac{32.042 \times 10^{-3}\text{kg}}{6.02214 \times 10^{23}}\right) \times 1.381 \times 10^{-23}\text{J} \cdot \text{K}^{-1} \times T}\right]^{\frac{3}{2}} - \ln\left[\frac{1000}{(41.30 \times 10^{-10}\text{m})^3}\right] + \frac{5}{2} = 3.3764 + \frac{3}{2}\ln T.$$

(45)

The rotational partition function of methanol is relatively difficult to construct since it requires an actual measurement of the molecular data in the mm to sub-mm wavelength regime. Fortunately, we found the experimental observable giving $\bar{B}$ = 4.25730, 0.82338, and 0.79273 cm$^{-1}$.[174] Converting $\bar{B}$ to the characteristic rotational temperatures using

$$\Theta_R = \frac{hc\bar{B}}{k_B} = 1.439 \times \bar{B}$$

(46)

results in $\Theta_R = 6.125, 1.185, 1.140$ K. Substituting these $\Theta_R$ values in Eq. (24) gives the final expression for the rigid rotor

$$s_{\text{rot}}^{(\text{id})} = -0.08295 + \frac{3}{2}\ln T.$$

(47)

Altogether, we finally arrive at the excess entropy expression for methanol

$$s_{ex}^{\text{FG}} = s^{\text{FG}} - \left(-0.08295 + \frac{3}{2}\ln T\right) - \left(3.3764 + \frac{3}{2}\ln T\right).$$

(48)

Figure 5(a) delineates the scaling relationship in the FG methanol. Remarkably, one obtains a linear relationship in the methanol system as well, given by

$$\ln D_{\text{FG}}^* = 0.65 \times s_{ex}^{\text{FG}} + 0.62.$$

(49)

As discussed above, we note differences in an exponent between $\alpha_{\text{H}_2\text{O}}^{\text{FG}}$ and $\alpha_{\text{MeOH}}^{\text{FG}}$ due to different rotational and translational contributions. However, as long as $\alpha_{\text{MeOH}}^{\text{FG}}$ is close to $\alpha_{\text{MeOH}}^{\text{CG}}$, our universality hypothesis remains valid.

To assess $\alpha_{\text{MeOH}}^{\text{CG}}$, we apply a similar procedure to CG methanol by only accounting for the translational degrees of freedom

$$s_{\text{trn}}^{(\text{id})}|_{\text{CG}} = \frac{S_{\text{trn}}^{(\text{id})}}{Nk_B}\Big|_{\text{CG}} = 3.3764 + \frac{3}{2}\ln T.$$

(50)

An overall trend observed in the CG systems is illustrated in Fig. 5(b), and a slight discrepancy observed in 400 K can be understood from the atomistic trend depicted in Fig. 3. Nevertheless, we find linearity over a wide range of temperatures from 250 K to 400 K in the following form

$$\ln D^*_{CG} = 0.65 \times s_{ex}^{FG} - 0.84.$$

(51)

In other words, we conclude that here again $\alpha_{MeOH}^{FG} = \alpha_{MeOH}^{CG}$. In other words, we conclude here again $\alpha_{MeOH}^{FG} = \alpha_{MeOH}^{CG}$, as can be seen in Fig. 5(c). The universality of $\alpha$ for methanol was further tested by checking if the same scaling relationship will hold for different resolutions. Based on the two-site CG mapping studied in Ref. 175, a two-site CG methanol model was constructed by mapping $CH_3$- and OH- atoms to CG sites. In this CG model, the CG excess entropy must consider additional rotational and vibrational motions along the C-O bond, i.e., Eqs. (45) to (48). Interestingly, Fig. 9 in Appendix A clearly demonstrates that two-site CG methanol also follows the nearly identical scaling relationship as the single-site model, confirming our perspective on universality.

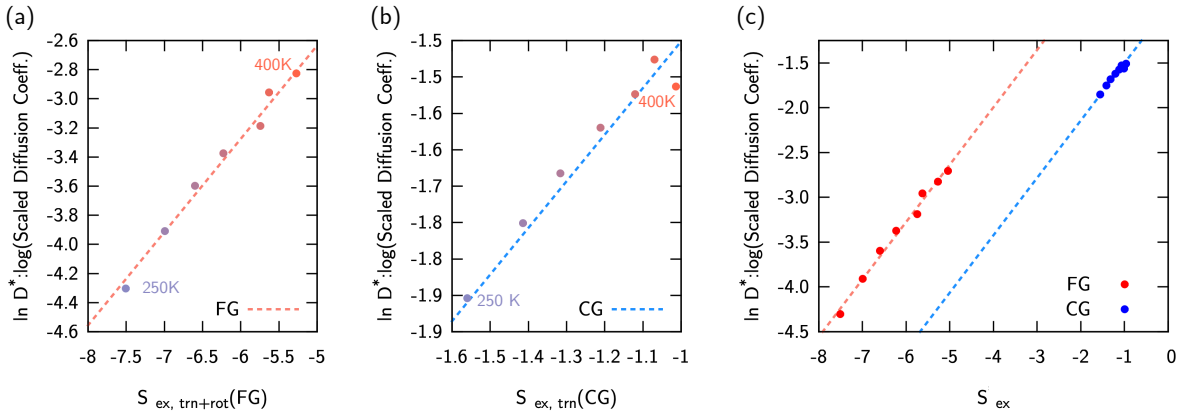

**Figure 5**: Examination of the Rosenfeld scaling in methanol. (a) FG scaling relationship over different temperatures from 250 K (bottom, left) to 400 K (top, right) fall into a linear scaling relation shown in Eq. (49) (dashed red line). (b) CG scaling relationship using the CG methanol model. Similarly, a linear excess entropy relationship is obtained (dashed blue line) from Eq. (50). (c) Overall comparison between the FG (red circle) and CG (blue circle) methanol models. Note that the slope (exponent) of the FG scaling (dashed red line) is identical to that of the CG scaling relation (dashed blue line), whilst only the y-intercept is different.

To conclude, we have discovered through an analysis of water and methanol that the excess entropy scaling relationship holds for both the FG and CG systems. More importantly, a universality relationship for simple liquids was elucidated for the first time that, given the same molecular system, the exponent (or slope in $\ln D^*$) from the scaling relationship remains invariant during the coarse-graining process. That is, $\alpha^{FG} = \alpha^{CG}$, indicating that the scaling relationship is intrinsically related to the underlying nature of the molecule, and thus it is still captured in the "bottom-up" CG models. Given the accelerated diffusion coefficient in the CG system, a relevant question is to then explicitly relate the CG diffusion coefficient to its FG analog in the scaled form.

**E. Relationship with Mapping Entropy**

From Eq. (5) and (6), the differences in diffusion coefficient between the FG and CG levels are due to two factors in the scaling relationship: $D_0$ and $S_{ex}$. The differences between the $S_{ex}^{FG}$ and $S_{ex}^{CG}$ terms can be clearly understood from the coarse-graining process. Since less important degrees of freedom are integrated out during the bottom-up coarse-graining process, the configurational entropy of the resultant CG model is reduced compared to the reference configurational entropy at the FG level. As discussed above, for the single-site CG model, this missing entropy, known as "mapping entropy," corresponds to the motions "beneath" the CG resolution: in this case rotation and vibration.[128] Figure 6 summarizes our understanding of the CG excess entropy scaling relationship with respect to the one obtained from the FG model. By decomposing the overall entropy into its underlying modes, it is evident that the motions observed in the CG dynamics are fully from translational motions $S^{CG} = S_{trn}^{CG}$, corresponding to the translational entropy at the FG resolution $S_{trn}^{CG}$. In other words, the differences in the excess entropy terms are primarily from the $S_{conf}^{FG}$ term and, importantly, indicating that one can adjust the CG entropy term with respect to the FG model to rescale the diffusion behavior.

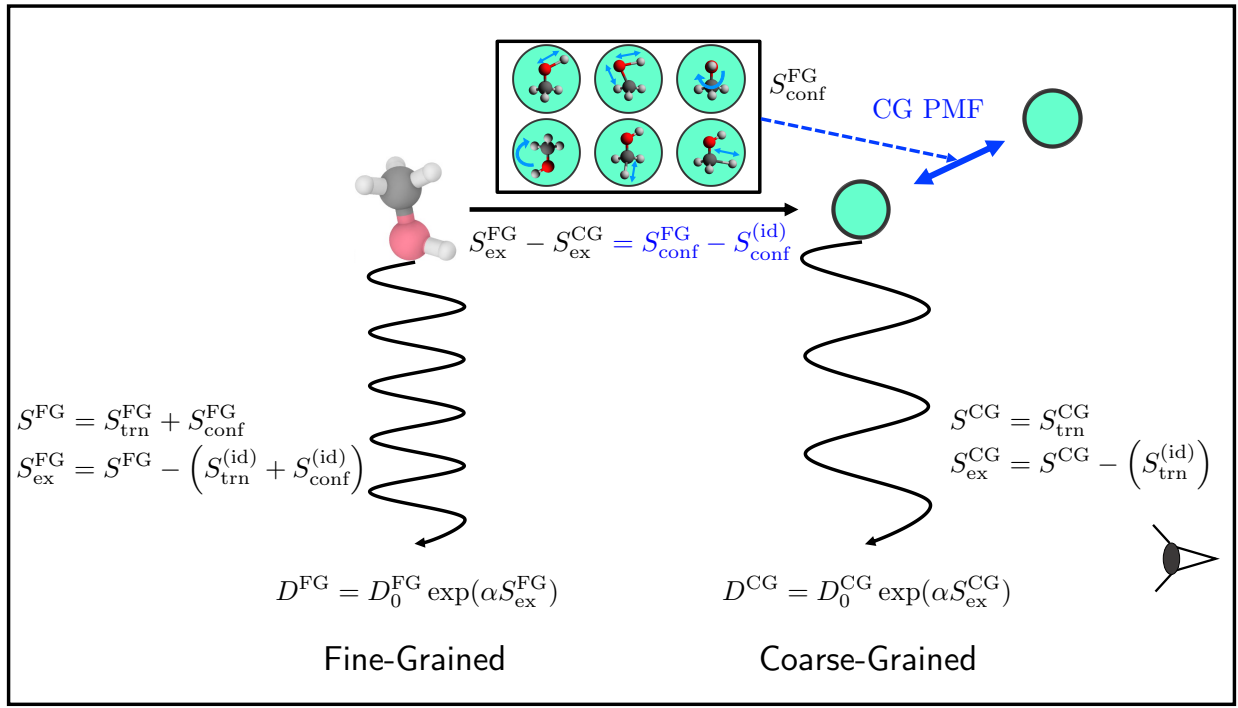

**Figure 6:** The present understanding of FG/CG dynamics in terms of excess entropy scaling. With an identical excess entropy scaling confirmed in this work, fast dynamics observed in the CG model can be understood from the mapping entropy $S_{conf}^{FG}$ and from different $D_0$ values.

**IV. Conclusions**

In this paper, we have elucidated the accelerated dynamical properties in CG fluid systems compared to the reference FG system by introducing an excess entropy scaling. While various studies have focused on the excess entropy in order to understand the FG dynamics, relatively less attention has been given to CG systems. However, based on the observation of missing motions in a given CG model and their corresponding contributions to the excess entropy, we assert that the excess entropy scaling can help to understand the speed-up of the CG dynamics. By combining the 2PT method and taking into consideration the orientational contributions, we were able to propose a comprehensive approach to computing the excess entropy for CG systems and the corresponding FG systems. Unlike other conventional approaches, we utilize the 2PT-based decomposition of molecular motions for the translational, vibrational, and rotational contributions, as this method has been shown to quantify the missing entropy during the coarse-graining process. By taking the ideal gas entropy from the Sackur-Tetrode equation for translation and using a rigid rotor entropy for rotation, we derived the excess entropy scaling behavior for both water and methanol at the atomistic (FG) level. By extending this approach to the CG systems, we observed the same scaling behavior exists with an identical scaling exponent for each molecular system. To our knowledge, this finding corroborates the universality and invariance of the excess entropy scaling under the coarse-graining process for liquid molecules studied in this work. Our finding also substantiates an earlier claim from Ref. 79 that proposed a connection between relative entropy differences and the dynamical behavior of liquids. Furthermore, the present theory proposes an alternative approach to accurately calculate the excess entropy for any molecular system and the corresponding CG model as long as one can construct ideal gas partition functions from known structural parameters. This requires a faithful construction of bottom-up CG models by considering the correct modal contribution to the excess entropy at the given resolution. Combining these pieces together, the present theory enables one to link the missing (or mapping) entropy with the faster CG dynamics followed by the identical scaling behavior as in the FG model.

As emphasized here, the present theory enables one to link the missing (or mapping) entropy with the faster CG dynamics. By correcting the missing entropy to CG systems, introducing the rotational entropies from the FG model would further lower the excess entropy, since $S_{ex}^{(\text{rot})} < 0$, resulting in slower CG dynamics. Hence, if the $D_0$ value for the FG model is known, it is possible to rescale the accelerated CG dynamics to the FG dynamics value by correcting the mapping entropy due to the coarse-graining following the formulation in Ref. 128 and adjusting $D_0^{\text{CG}}$ to $D_0^{\text{FG}}$. We further applied this idea to CG water systems in Fig. 7 below, where the recovered FG diffusion coefficients from CG dynamics $D_{\text{CG}}^{\text{corr}}$ show almost identical values as the original FG diffusion with an error of 4.5% for SPC/E, 2.1% for SPC/Fw, 5.7% for TIP4P/2005, and 2.4% for TIP4P/Ice. We note that a similar approach was recently suggested by Rondina et al. for rescaling dynamics at different CG resolutions of Lennard-Jones chains of polymers using the Dzugutov scaling.[80]

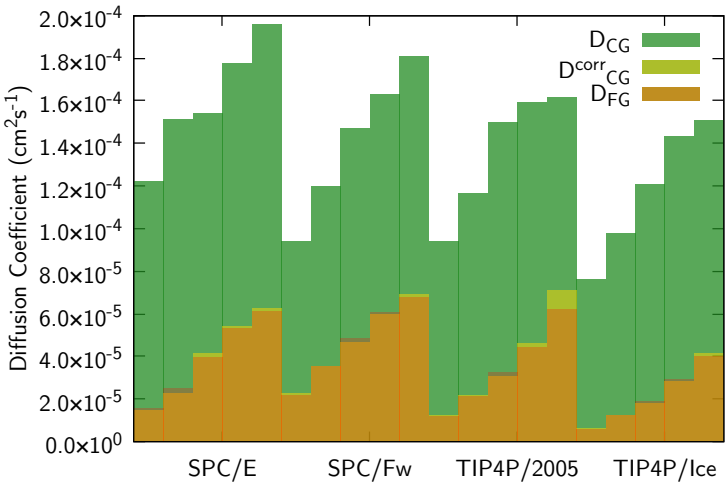

**Figure 7:** Corrected diffusion coefficients of water (yellow bars) from CG (green bars) given the FG scaling intercept $D_0^{\text{FG}}$. Note that the rescaled CG diffusion coefficients can reproduce the original FG diffusion coefficients (orange bars). Four force fields from 280 K to 360 K at 20 K intervals are considered.

Nevertheless, for complex molecules governed by the Rosenfeld scaling, the *ad hoc* nature of $D_0$ and alpha values are a current bottleneck to bridge dynamics between the FG and CG dynamics. To elaborate more on this viewpoint, Fig. 8 demonstrates that the corrected CG diffusion coefficient and reference FG diffusion coefficient simply based on the excess entropy differences do not completely agree with each other due to the difference in the $D_0$ terms. Despite having the same scaling exponent, we see differences in the $D_0$ values (which we think of as an "entropy-free" diffusion coefficient) between the FG and CG models. Thus, a better understanding of the physical meaning and origin of the $D_0$ term, as well as their excess entropy difference, is required.

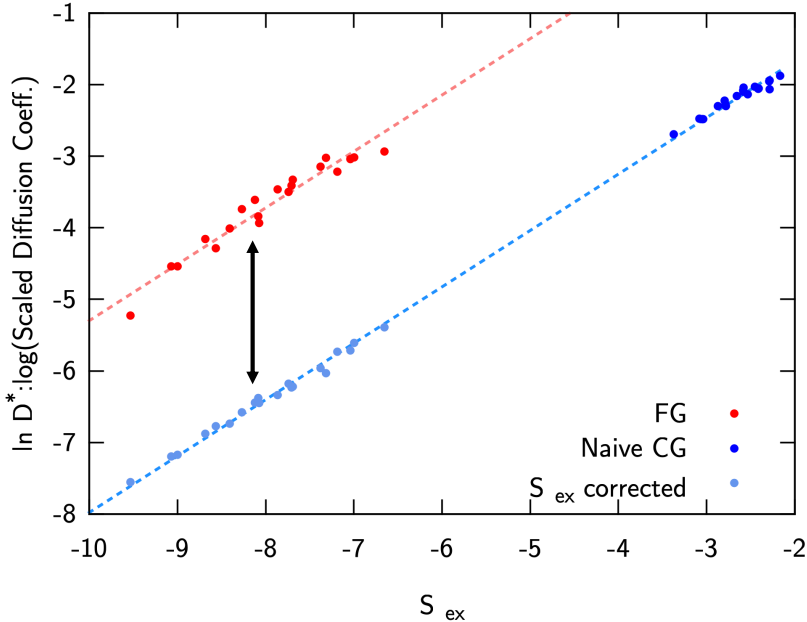

**Figure 8:** Excess entropy scaling (dashed lines) of water for FG (red dot) and CG systems (blue dot) and after adjusting the mapping entropy to the CG model (sky blue dot).

Nevertheless, the phenomenological foundations of the scaling behavior limit further theoretical analysis of the $D_0$ term for both the FG and CG systems. Even though the FG dynamics involve complicated coupled motions with different time scales, the single-site CG resolution can integrate these extraneous motions out, leaving the translational motions only. Thus, in order to systematically better understand the meaning of $D_0$ and to develop approaches for its calculation, it may be physically sound to approximate the reduced CG dynamics as hard sphere dynamics. This approach is pursued further in Paper II of this series with the goal of providing a more complete description of the CG model diffusion based on the universal scaling law between the CG and FG systems of fluids as uncovered in this paper.[176]

**ACKNOWLEDGMENTS**
This material is based upon work supported by the National Science Foundation (NSF Grant CHE-2102677). Simulations were performed using computing resources provided by the University of Chicago Research Computing Center (RCC). J.J. acknowledges the Harper Dissertation Fellowship from the University of Chicago and insightful discussions with Professor Jeppe C.

Dyre. K.S.S acknowledges the support of the University of Chicago and the Pritzker School of Molecular Engineering during his sabbatical stay where this work was initiated.

## DATA AVAILABILITY
The data that support the findings of this work are available from the corresponding author upon request.

## APPENDIX
### APPENDIX A: Two-site CG Methanol Model
The two-site CG model for methanol was constructed by following the same procedures described in Ref. 175. First, the CG sites were determined by mapping $CH_3$- and $OH$- atoms to the center-of-mass of each group. Parametrization settings for non-bonded interactions were kept identical to the single-site case, while the bonded interaction was described by the $4^{th}$ order B-splines with smaller spacing (0.025 Å). After parametrizing the two-site CG methanol model from MS-CG, the same simulation protocol was used to run CG simulations at different temperatures as Fig. 5. The excess entropy scaling was accessed by computing the additional rotational and vibrational motions remained at the two-site CG level. In Fig. 9, we observe that the fitted slope of the two-site CG model differs by a value of 0.009 compared to the single-site CG scaling relationship. Hence, we believe that the two-site CG model also follows the nearly identical scaling exponent as the single-site CG model, indicating that the scaling relationship is invariant for methanol under different resolutions.

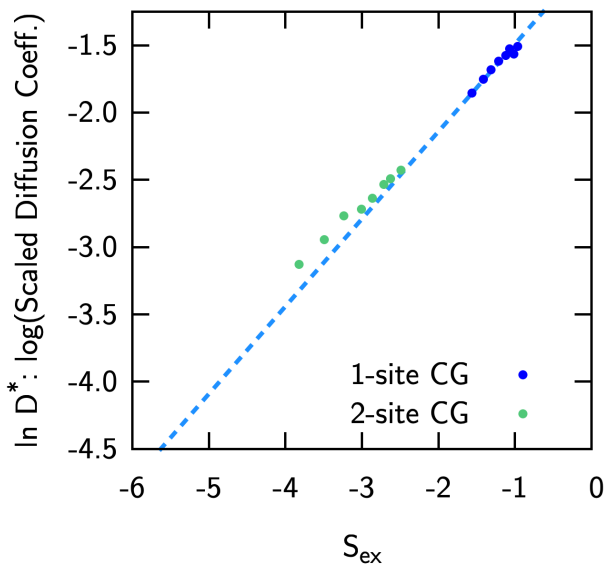

**Figure 9:** Rosenfeld scaling for CG methanol at different resolutions. Results from the 1-site CG model (blue circles) and a linear relationship from Fig. 5 (dashed blue line) were directly on top of the results from the 2-site CG model (green circle). Differences between the fitted slopes and $y$-intercept for two CG resolutions are obtained as 0.009 and 0.03, respectively, and are deemed negligible.

## REFERENCES
1. F. Müller-Plathe, ChemPhysChem **3** (9), 754-769 (2002).